\documentclass[12pt]{iopart}
\usepackage{cite}
\usepackage{graphicx}
\usepackage{subfig}
\usepackage{amssymb}
\begin{document}

\letter{On conversion of high-frequency soliton solutions to a (1+1)-dimensional nonlinear evolution equation}

\author{\textsc{Kuetche Kamgang Victor}$^{1}$, \textsc{Bouetou Bouetou Thomas}$^{2,3}$ and \textsc{Kofane Timoleon Crepin}$^{1,3}$}

\address{$^{1}$ Department of Physics, Faculty of Science,
University of Yaounde I, P.O. Box. 812
Cameroon \\
$^{2}$ Ecole Nationale Sup$\acute{e}$rieure Polytechnique,
University of Yaounde I. P.O. Box. 8390, Cameroon\\
$^{3}$ The Abdus Salam International Centre for Theoretical
Physics. P.O. Box 586, Strada Costiera, II-34014, Trieste, Italy}
\ead{vkuetche@yahoo.fr, tbouetou@yahoo.fr and tckofane@yahoo.com}
\begin{abstract}
We derive a (1+1)-dimensional nonlinear evolution equation (NLE)
which may model the propagation of high-frequency perturbations in
a relaxing medium. As a result, this equation may possess three
typical solutions depending on a dissipative parameter.
\end{abstract}
 \vspace{2pc}

Nonlinear dynamics may be a topic of interest in various fields of
science and engineering. A lot of problems arising in science and
engineering may be modelled as a dynamical system. As an
illustration, there is the Van der Pol equation \cite{Van} given
by
\begin{eqnarray}
\frac{d^{2}u}{dt^{2}}-\mu(1-u^{2})\frac{du}{dt}+\omega_{0}^{2}u=0,\label{eq1}
\end{eqnarray}
found in nonlinear circuit theory \cite{Guk}. The quantity $u$ is
a physical observable depending on the time $t$; $\mu$ and
$\omega_{0}$ are constants. There is also the Van der Pol-Duffing
equation \cite{Ued} given by
\begin{eqnarray}
\frac{d^{2}u}{dt^{2}}-\mu(1-u^{2})\frac{du}{dt}+\omega_{0}^{2}u+\nu
u^{3}=0,\label{eq2}
\end{eqnarray}
which may model the optical bistability in a dispersive medium
\cite{Kao}. The additional quantity $\nu$ is a cubic parameter.

Besides, nonlinear phenomena may be also described by nonlinear
partial differential (NLPD) equations such as the well-known
Vakhnenko equation \cite{Vak1} given by
\begin{eqnarray}
u_{xt}+\frac{1}{2}(u^{2})_{xx}+u=0,\label{eq3}
\end{eqnarray}
arising in relaxing media as a model equation of propagation of
high-frequency perturbations. Subscripts denote partial
differentiation with respect to time $t$ and space $x$. Equation
\eref{eq3} has been subject to many investigations
(\cite{Vak2,Vak3,Mor1,Mor2} and references therein) in recent
years. One typical class of solutions to NLPD equations are the
so-called solitons arising as the result of balance between
nonlinear and dispersion effects. Higher order solitons may be
also found in higher order NLPD equations. One important question
that may be pointed out is whether higher order solitons may
survive in higher order Vakhnenko equation \cite{Vak1}.

In the present letter, we consider a barotropic medium
$p=p(\rho,\lambda)$ under relaxation. The quantities $p$ and
$\rho$ denote pressure and mass density, respectively, and
$\lambda$ is an additional parameter. Then, we derive a novel
(1+1)-dimensional NLE model equation. We discuss the different
soliton solutions to this (1+1)-dimensional NLE equation.

Recently, Vakhnenko \cite{Vak1} has derived a dynamic state
equation using an expansion of the specific volume $V$ as power
series of the small perturbations $p'\ll p_{0}$ with accuracy
$o(p'^{2})$. The quantity $p_{0}$ is the pressure related to the
unperturbed state. Performing this expansion to accuracy
$o(p'^{3})$, the following dynamic state equation may be found
\begin{eqnarray}
\fl\tau\left(p'_{xx}-p'_{tt}/v_{f}^{2}+\alpha_{f}\left(p'^{2}\right)_{tt}+a_{f}\left(p'^{3}\right)_{tt}\right)_{t}+p'_{xx}-p'_{tt}/v_{e}^{2}+\alpha_{e}\left(p'^{2}\right)_{tt}+a_{e}\left(p'^{3}\right)_{tt}=0,\label{eq4}
\end{eqnarray}
where $\tau$ is the relaxation time,
$\alpha_{i}=\frac{1}{2V_{0}^{2}}\frac{d^{2}V_{i}}{dp^{2}}|_{p=p_{0}}$,
$a_{i}=\frac{1}{6V_{0}^{2}}\frac{d^{3}V_{i}}{dp^{3}}|_{p=p_{0}}$
and $v_{i}$ may stand for velocities of the relaxation processes
defined as $v_{i}^{2}=\frac{dp_{i}}{d\rho}$. For high-frequency
perturbations, that is $i=f$ and $p'=p'_{f}=p'(\rho,1)$ and for
low-frequency perturbations, that is $i=e$ and
$p'=p'_{e}=p'(\rho,0)$. Following the ref. \cite{Vak1}, equation
\eref{eq4} may be analyzed by means of the multiscale method
\cite{Nay,Nit} by introducing a small parameter
$\epsilon=\tau\omega$ where $\omega$ is the frequency of the wave
perturbation.

Using a dispersion relation of the form
$\omega=v_{e}k+j\beta_{e}k^{2}-\gamma_{e}k^{3}$ where $j^{2}=-1$
for the linearized equation \eref{eq4}, in the case of
low-frequency perturbations, that is $\tau\omega\ll 1$, a
(1+1)-dimensional NLE equation may be derived as follows
\begin{eqnarray}
p'_{t}+v_{e}p'_{x}+\alpha_{e}v_{e}^{3}\left(p'^{2}\right)_{xx}+a_{e}v_{e}^{3}\left(p'^{3}\right)_{x}-\beta_{e}p'_{xx}+\gamma_{e}p'_{xxx}=0,\label{eq5}
\end{eqnarray}
with
\begin{eqnarray}
\beta_{e}=\frac{v_{e}^{2}\tau}{2v^{2}_{f}}\left(v^{2}_{f}-v^{2}_{e}\right),\quad
\gamma_{e}=\frac{v_{e}^{3}\tau^{2}}{8v^{4}_{f}}\left(v^{2}_{f}-v^{2}_{e}\right)\left(v^{2}_{f}-5v^{2}_{e}\right).
\label{eq6}
\end{eqnarray}
The nonlinear terms have been reconstructed in agreement with the
initial equation. This equation \eref{eq5} may be viewed as a
modified Korteweg-de Vries-Burgers (mKdVB) equation. Recently, Fu
et al. \cite{Fu} have studied the specific case $\alpha_{e}=0$ and
$\beta_{e}=0$ (mKdV equation) by constructing breather lattice
solutions. Similar procedure may be discussed for equation
\eref{eq5} in order to find out other kind of breather lattice
solutions. We may actually think that equation \eref{eq5} may
deserve further interests from the viewpoint of investigation of
propagation of localized and periodic waves.

In the case of high-frequency perturbations, that is
$\tau\omega\gg 1$, performing the dispersion relation to
$v_{f}^{-2}\omega=k^{2}+j\beta_{f}k-\gamma_{f}$, in agreement with
the initial equation, one may get the following (1+1)-dimensional
NLE equation
\begin{eqnarray}
p'_{xx}-v_{f}^{-2}p'_{tt}+\alpha_{f}v_{f}^{2}\left(p'^{2}\right)_{xx}+a_{f}v_{f}^{2}\left(p'^{3}\right)_{xx}+\beta_{f}p'_{x}+\gamma_{f}p'=0,\label{eq7}
\end{eqnarray}
where
\begin{eqnarray}
\beta_{f}=\frac{v_{f}^{2}-v_{e}^{2}}{\tau v_{e}^{2}v_{f}},\quad
\gamma_{f}=\frac{v_{f}^{4}-v_{e}^{4}}{2\tau^{2}v_{e}^{4}v^{2}_{f}},
\label{eq8}
\end{eqnarray}
standing for dissipative and dispersive parameters. In order to
investigate the equation \eref{eq7}, it seems useful to consider
the following accuracy
\begin{eqnarray}
\partial_{x}^{2}-v_{f}^{-2}\partial_{t}^{2}\approx 2\partial_{x}\left(\partial_{x}+v_{f}^{-1}\partial_{t}\right).\label{eq9}
\end{eqnarray}
We consider two interesting cases: $\alpha_{f}=0$ and
$\alpha_{f}\neq 0$.
\begin{enumerate}
    \item First case: $\alpha_{f}=0$.

    Equation \eref{eq7} may be reduced to
    \begin{eqnarray}
    u_{y\eta}-\frac{1}{6}(u^{3})_{yy}+\alpha u_{y}-u=0,\label{eq9}
    \end{eqnarray}
    up to the following transformations
    \begin{eqnarray}
    \fl y=\sqrt{\frac{\gamma_{f}}{6}}\left(v_{f}t-x\right),\quad \eta=\sqrt{\frac{3\gamma_{f}}{2}}v_{f}t,\quad \alpha=\beta_{f}\sqrt{\frac{1}{6\gamma_{f}}},\quad p'=\frac{u}{v_{f}\sqrt{a_{f}}}.\label{eq10}
    \end{eqnarray}
    Without dissipative $\alpha$-term, equation \eref{eq9} may be observed as the Sch$\ddot{a}$fer-Wayne short pulse (SWSP) equation \cite{Sch}
    which has been subject to many recent investigations \cite{Chu,Sak1,Sak2,Kue1,Kue2,Kue3,Par}. This SWSP equation may have a variant form given by
    \begin{eqnarray}
    u_{xt}+\frac{1}{6}(u^{3})_{xx}+u=0,\label{eqq9}
    \end{eqnarray}
    up to the transformations $x$$\rightarrow$$jx$,
    $y$$\rightarrow$$jy$, $t$$\rightarrow$$jt$ and
    $u$$\rightarrow$$ju$, $j^{2}=-1$.
    Further interests ought to be paid to equation \eref{eq9}
    which may have many applications in soliton theory and
    nonlinear optics. In particular, its extension to a
    complex-valued SWSP equation \cite{Kue3} with dissipative $\alpha$-terms may be
    worth investigating alongside the effect of the dissipative
    parameter $\alpha$ on the different solutions. Indeed,
    extending $u$ to a complex-valued quantity $Q$ \cite{Kue3} in order to get the following equation
    \begin{eqnarray}
    Q_{y\eta}-\frac{1}{2}\left(|Q|^{2}Q_{y}\right)_{y}+\alpha Q_{y}-Q=0,\label{eqq10}
    \end{eqnarray}
    equation \eref{eqq10} may be transformed into the following system
    \begin{equation}
    \eqalign{ Q^{r}_{\sigma\sigma}-Q^{r}_{\tau\tau}=(Z_{\sigma}+Z_{\tau})Q^{r}-\alpha (Q^{r}_{\sigma}+Q^{r}_{\tau}),\cr
    Q^{im}_{\sigma\sigma}-Q^{im}_{\tau\tau}=(Z_{\sigma}+Z_{\tau})Q^{im}-\alpha (Q^{im}_{\sigma}+Q^{im}_{\tau}),\cr
    Z_{\sigma\sigma}-Z_{\tau\tau}=-Q^{r}(Q^{r}_{\sigma}+Q^{r}_{\tau})-Q^{im}(Q^{im}_{\sigma}+Q^{im}_{\tau}),}\label{eqq11}
    \end{equation}
    up to the following transformations
    \begin{eqnarray}
    y=-\left(X+\frac{1}{2}\int_{-\infty}^{T}QQ^{\star}dT'\right)+\mu,\quad \eta=T,\label{eqq12}
    \end{eqnarray}
    where $\mu$ is an arbitrary constant, $T=\frac{1}{2}(\sigma-\tau)$ and $X=-\frac{1}{2}(\sigma+\tau)$.
    Looking for soliton solutions with the boundary conditions $|Q|\rightarrow 0$, $Z\rightarrow \sigma/2$ as
    $|\sigma|\rightarrow \infty$, equation \eref{eqq11} may be bilinearized as in ref \cite{Kue3} according to
    the Hirota's method \cite{Hir1,Hir2}, and soliton solutions to equation \eref{eqq11} may be easily derived.
    Thus, deriving the dispersion relation which may obviously be expressed in terms of the dissipative parameter $\alpha$,
    it is then possible to discuss the soliton solutions of equation \eref{eqq10} with respect to $\alpha$.
    As a result, a soliton solution $Q$ may be given by
    \begin{eqnarray}
    Q=Asech(\vartheta^{r})\exp(\imath \vartheta^{im}),\label{eqq13}
    \end{eqnarray}
    where $\vartheta=k\sigma-\omega\tau+\vartheta_{0}$, $\vartheta_{0}$ being a contant parameter, and
    \begin{eqnarray}
    A=4(k^{r}+\omega^{r}).\label{eqq14}
    \end{eqnarray}
    The physical complex-valued quantities $k$ and $\omega$
    standing for wave number and angular frequency, respectively,
    may satisfy the following dispersion equation
    \begin{eqnarray}
    k^{2}-\omega^{2}+\alpha(k-\omega)-1=0,\label{eqq15}
    \end{eqnarray}
    from which $k$ and $\omega$ may be expressed in terms of the dissipative
    parameter $\alpha$. Thus, discussing the soliton solutions
    expressed in terms of $Q^{r}$ and $Q^{im}$ vs $y$, one may expect to find loop-, cusp- and hump- shaped
    solitons. For some convenience, we do not go further with
    these developments. We shall focus our attention to a further
    novel equation below that may be also of great interests.

    \item Second case: $\alpha_{f}\neq 0$.

    Equation \eref{eq7} may be reduced to
    \begin{eqnarray}
    \partial_{y}\left(\partial_{\eta}+u\partial_{y}+\frac{u^{2}}{2}\partial_{y}\right)u+\alpha u_{y}+u=0,\label{eq11}
    \end{eqnarray}
    up to the following transformations
    \begin{eqnarray}
    \fl y=\frac{1}{\alpha_{f}}\sqrt{\frac{3a_{f}\gamma_{f}}{2}}\left(v_{f}^{-1}x-t\right),\quad \eta=\sqrt{\frac{\gamma_{f}}{6a_{f}}}\alpha_{f}v_{f}^{2}t,\quad \alpha=\frac{\beta_{f}}{\alpha_{f}v_{f}}\sqrt{\frac{3a_{f}}{2\gamma_{f}}},\quad p'=\frac{\alpha_{f}u}{3a_{f}}.\label{eq12}
    \end{eqnarray}
    Without the dissipative term and
    $\partial_{y}\left(\frac{u^{2}}{2}\partial_{y}u\right)$-term, equation
    \eref{eq11} may be reduced to the well-known Vakhnenko
    equation \cite{Vak1}. Without the dissipative term and $\partial_{y}\left(u\partial_{y}u\right)$-term,
    equation \eref{eq11} may be reduced to \eref{eqq9}.
    Performing variable transformations, we introduce new
    independent variables $\xi$ and $\zeta$ as follows
    \begin{eqnarray}
    y=\zeta+\int_{-\infty}^{\xi}\left(u+\frac{1}{2}u^{2}\right)d\xi'+y_{0},\quad \eta=\xi,\label{eq13}
    \end{eqnarray}
    where $y_{0}$ is an arbitrary constant. Then, equation \eref{eq11} is
    reduced to
    \begin{eqnarray}
    u_{\xi\zeta}+\alpha u_{\zeta}+\varphi u=0,\label{eq14}
    \end{eqnarray}
    where
    \begin{eqnarray}
    \varphi=1+\int_{-\infty}^{\xi}u_{\zeta}\left(1+u\right)d\xi'.\label{eq15}
    \end{eqnarray}
    Defining another independent variables $\sigma$ and $\tau$ as follows
    \begin{eqnarray}
    \xi=\frac{1}{2}(\sigma-\tau),\quad \zeta=-\frac{1}{2}(\sigma+\tau),\label{eq16}
    \end{eqnarray}
    equation \eref{eq14} is transformed to
    \begin{eqnarray}
    u_{\sigma\sigma}-u_{\tau\tau}=\varphi u-\alpha (u_{\sigma}+u_{\tau}).\label{eq17}
    \end{eqnarray}
    Moreover, using the ansatz
    \begin{eqnarray}
    \varphi=-Z_{\zeta}=(Z_{\sigma}+Z_{\tau}),\label{eq18}
    \end{eqnarray}
    we get the following coupled equations
    \begin{equation}
    \eqalign{ u_{\sigma\sigma}-u_{\tau\tau}=(Z_{\sigma}+Z_{\tau}) u-\alpha (u_{\sigma}+u_{\tau}),\cr
    Z_{\sigma\sigma}-Z_{\tau\tau}=-u(u_{\sigma}+u_{\tau})-(u_{\sigma}+u_{\tau}).}\label{eq19}
    \end{equation}
    This system may be closely related to that described by Kakuhata and Konno
    \cite{Kak} while investigating the loop soliton solutions of
    string interacting with external field. Thus, the other
    physical meaning of equation
    \eref{eq11} is pointed out. This
    may be useful in constructing the soliton solutions to equation
    \eref{eq11}. Thus, in order to find a soliton solution, we
    consider the following boundary conditions
    \begin{eqnarray}
    u\rightarrow 0,\quad Z\rightarrow \sigma/2,\quad \verb"as"\quad
    \sigma\rightarrow -\infty.\label{eq20}
    \end{eqnarray}
    We may consider the following settings \cite{Kak}
    \begin{eqnarray}
    u=\frac{G}{F},\quad Z=\frac{1}{2}(\sigma+\tau)+2(\partial_{\tau}-\partial_{\sigma})\ln F.\label{eq21}
    \end{eqnarray}
    Equation \eref{eq19} is then bilinearized as follows
    \begin{equation}
    \eqalign{\left(D_{\sigma}^{2}-D_{\tau}^{2}+\alpha\left(D_{\sigma}+D_{\tau}\right)^{2}-1\right)(F\cdot G)=
    0,\cr
    \left(D_{\sigma}-D_{\tau}\right)^{2}(F\cdot F)-\frac{1}{2}\left(G^{2}+2GF\right)=0,}\label{eq22}
    \end{equation}
    where $D_{\sigma}$ and $D_{\tau}$ denote Hirota operators \cite{Hir1,Hir2}.
    Expanding $F$ and $G$ in a suitable formal power series, a
    soliton solution to equation \eref{eq19} is given by
    \begin{eqnarray}
    \fl u=4(\omega+k)^{2}\left[\tanh(\theta)+1\right],\quad Z=\frac{1}{2}(\sigma+\tau)-2(\omega+k)\left[\tanh(\theta)+1\right].\label{eq23}
    \end{eqnarray}
    where $\theta=k\sigma-\omega\tau+\theta_{0}$, $\theta_{0}$
    being an arbitrary constant. The dispersion relation is given
    by
    \begin{figure}
    \begin{center}
    \subfloat[$\alpha=0.351648275547$]{\includegraphics[width=6.5cm]{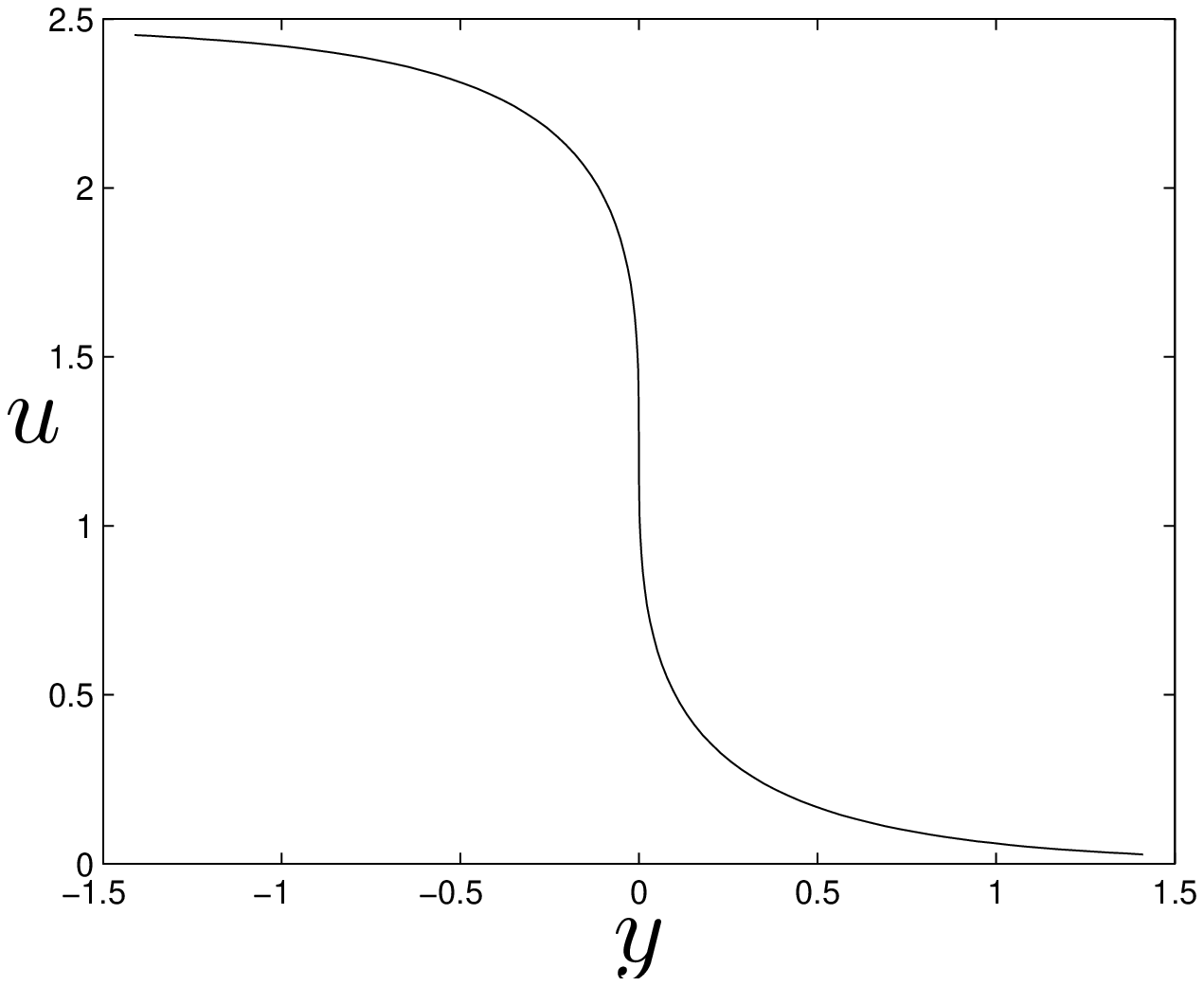}}\label{fig1a}
    \subfloat[$\alpha=0.351648275547$]{\includegraphics[width=6.5cm]{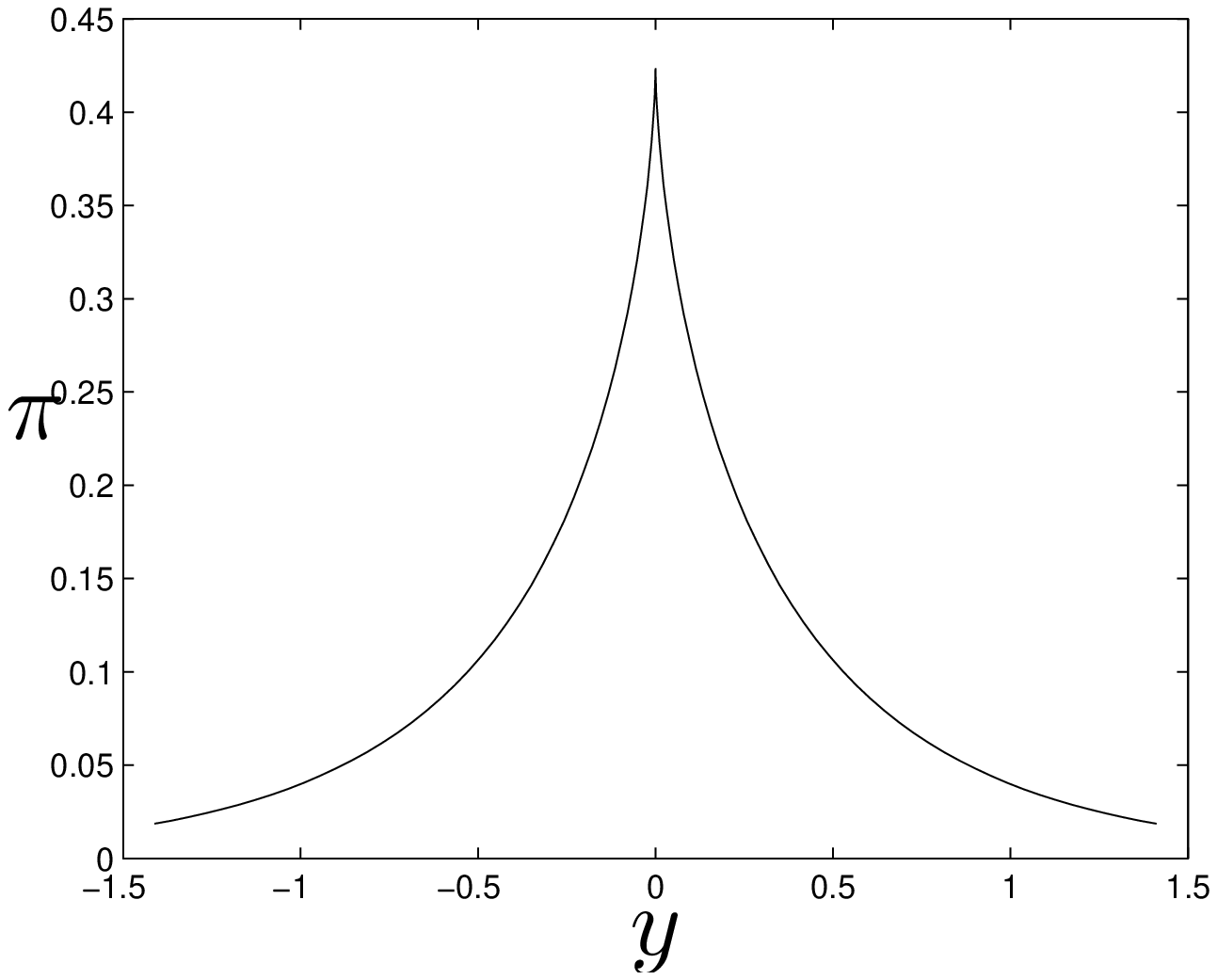}}\label{fig1b}\\
    \subfloat[$\alpha=0.1$]{\includegraphics[width=6.5cm]{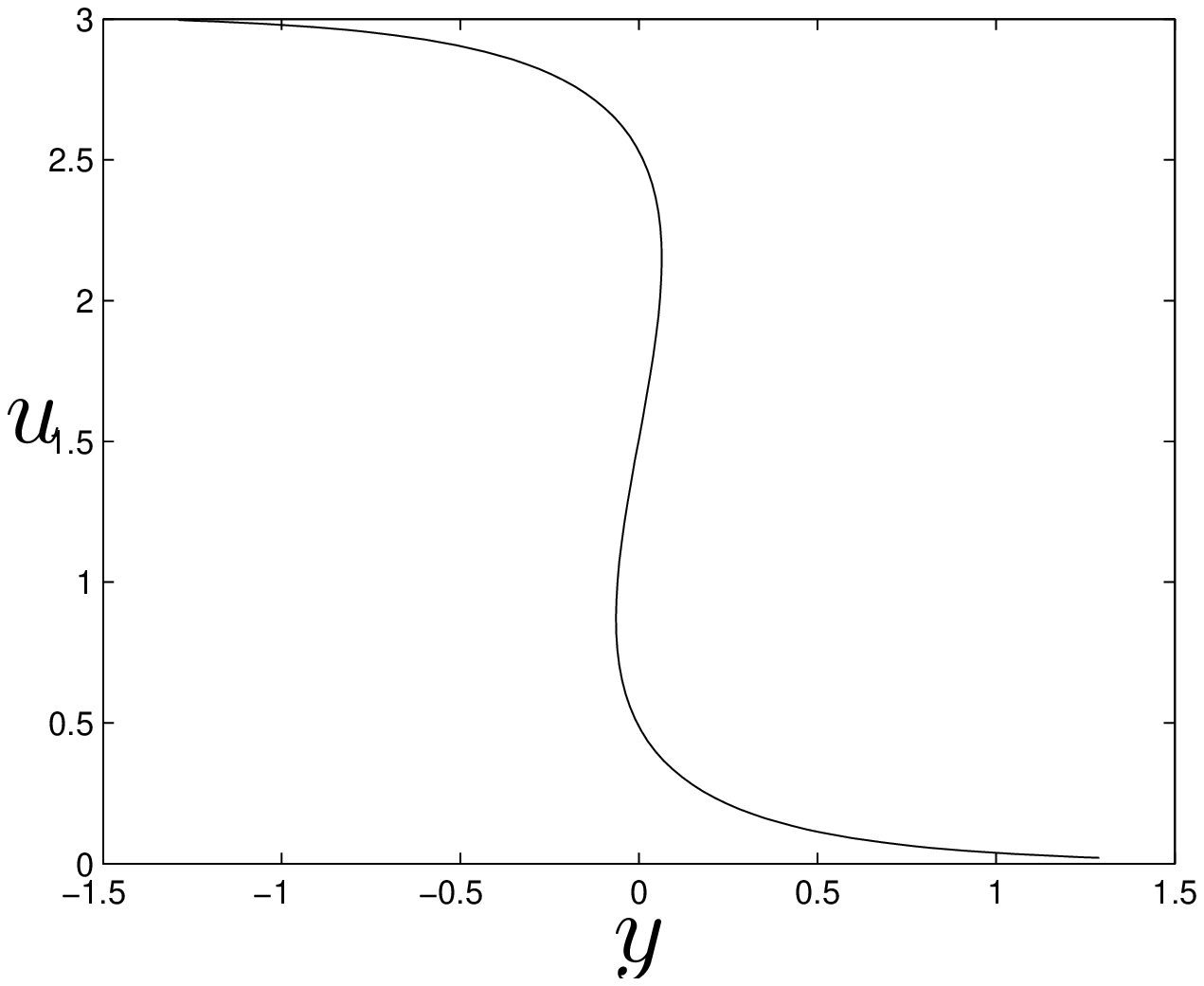}}
    \subfloat[$\alpha=0.1$]{\includegraphics[width=6.5cm]{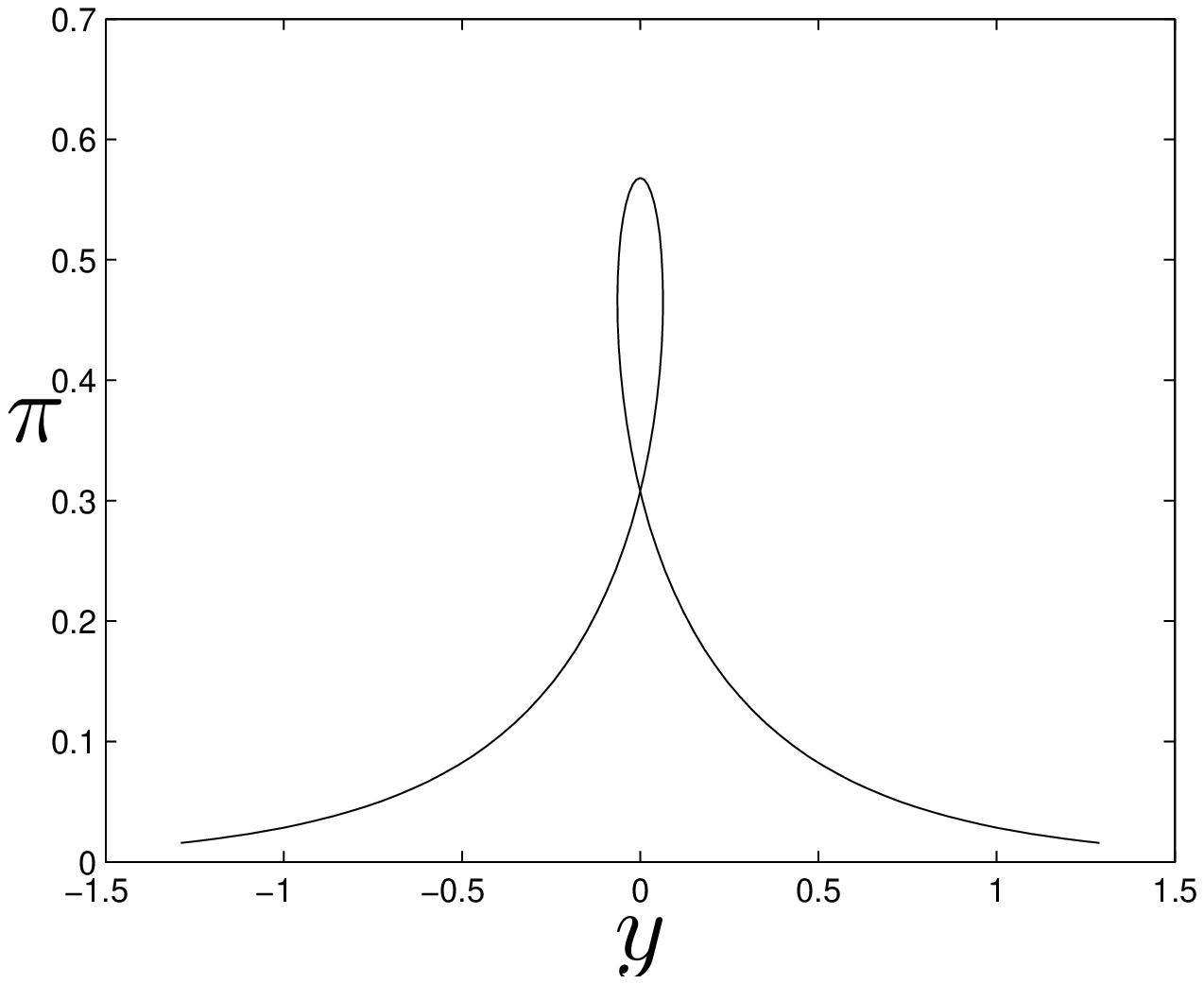}}\\
    \subfloat[$\alpha=0.8$]{\includegraphics[width=6.5cm]{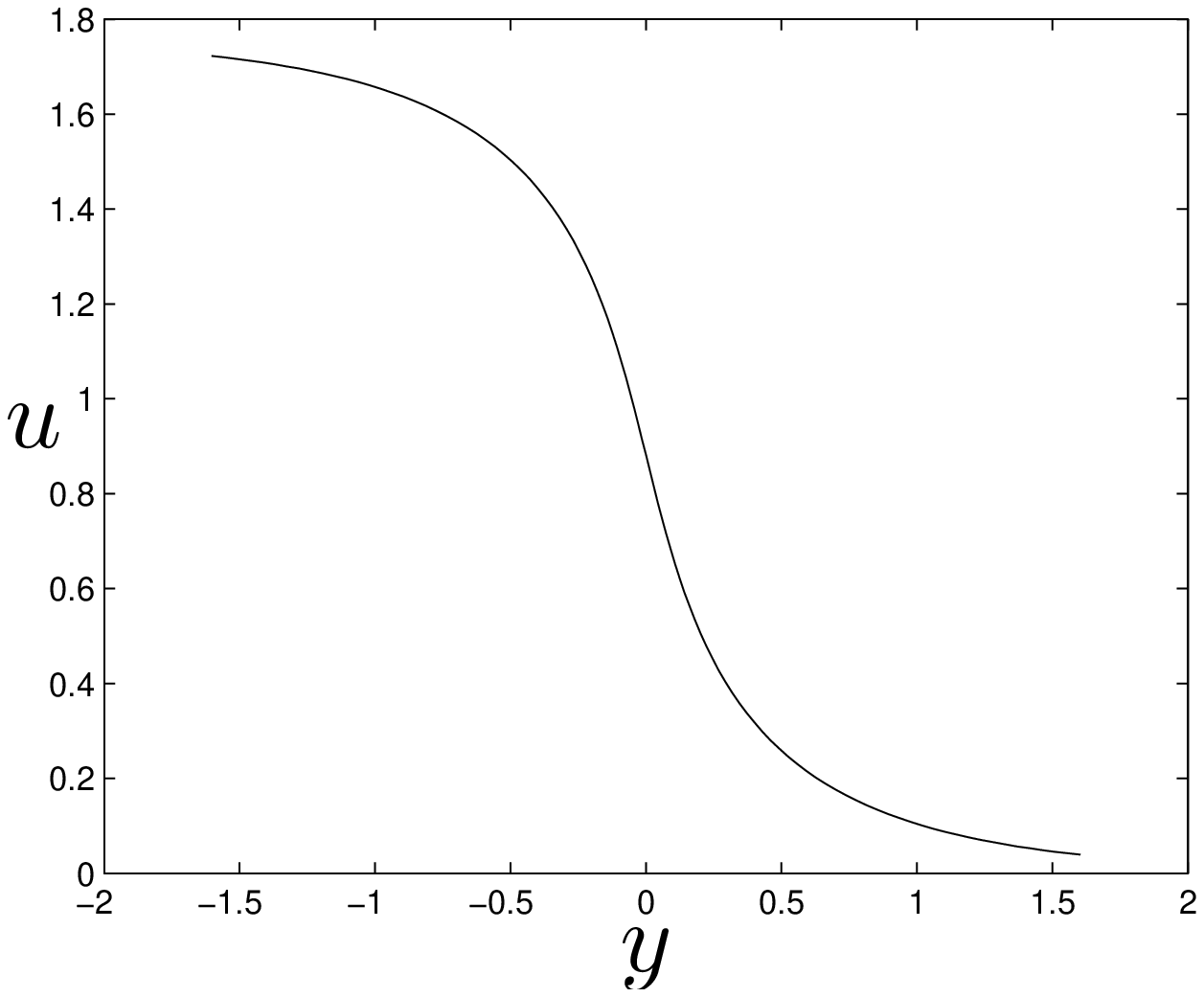}}
    \subfloat[$\alpha=0.8$]{\includegraphics[width=6.5cm]{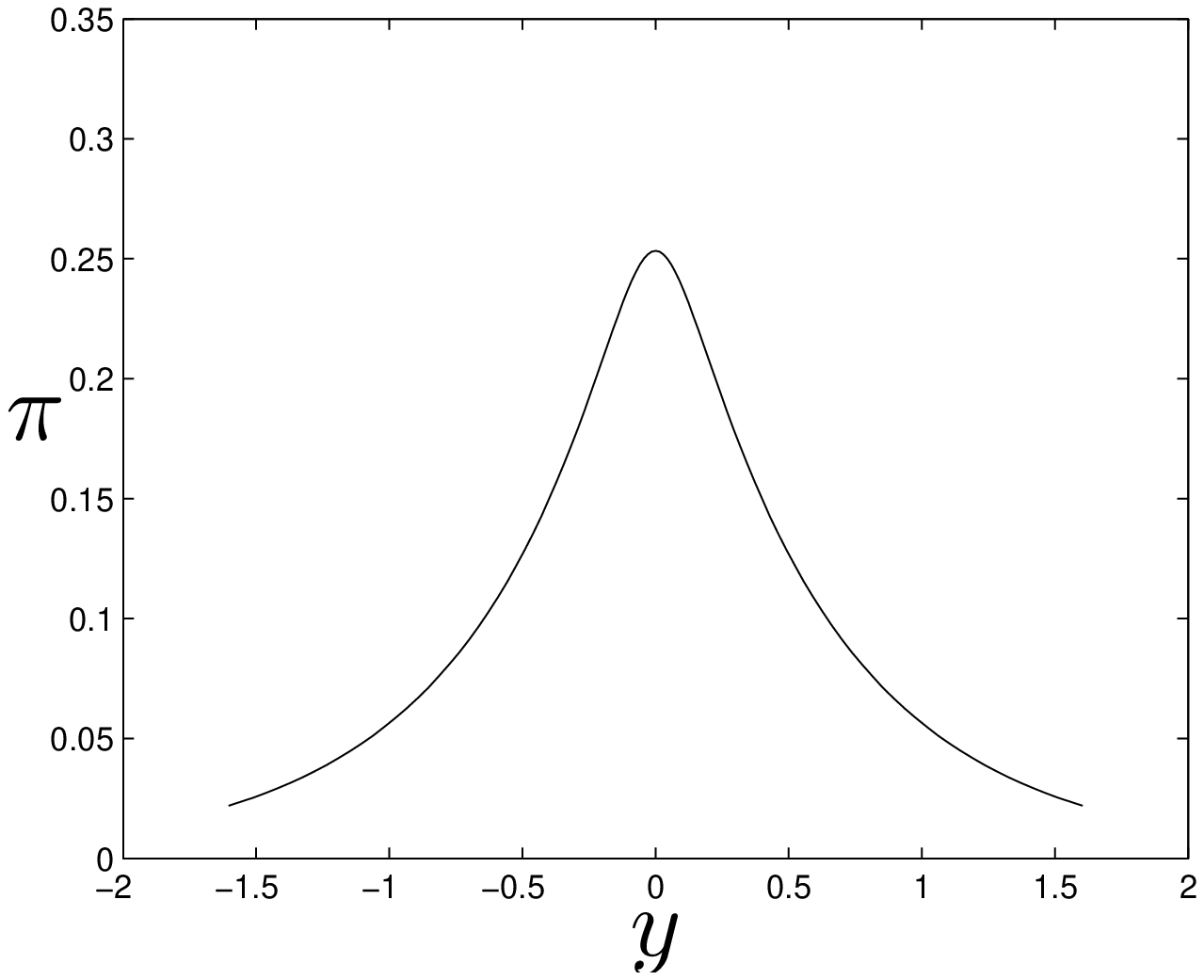}}\\
    \caption{Shape $u$ and corresponding momentum $\pi$ of the soliton.}\label{fig}
    \end{center}
\end{figure}
    \begin{eqnarray}
    4(k^{2}-\omega^{2})+2\alpha(k-\omega)-1=0,\label{eq24}
    \end{eqnarray}
    which may lead to the following solution ($k>0$)
    \begin{eqnarray}
    k=\frac{1}{\alpha(1-v)+\sqrt{\alpha^{2}(1-v)^{2}+4(1-v^{2})}},\quad \omega=kv,\label{eq25}
    \end{eqnarray}
    where $v$ is the velocity of the wave satisfying the condition
    $-1<v<1$. It seems worth noting here that from equations
    \eref{eq13}, \eref{eq15} and \eref{eq18}, one may find that
    $y=-Z+C$, $C$ being an arbitrary constant. In order to
    discuss the soliton solutions to equation
    \eref{eq11}, it is important to consider the following relation
    \begin{eqnarray}
    \partial_{\sigma}=\frac{1}{2}\left[1-4(\omega+k)k \verb"sech"^{2}(\theta)\right]\partial_{Z}.\label{eq26}
    \end{eqnarray}
    We may pay interest to the shape of the soliton $u$ and its
    momentum $\pi=u_{\sigma}+u_{\tau}$. As a result, it comes that
    \begin{itemize}
        \item for $\alpha=v\frac{\sqrt{1+v}}{1-v}$, $u_{Z}$ may never
        change sign but may be infinite at some 'singular' point, whereas
        $\pi_{Z}$ may change sign once and may be infinite at the same singular point. Thus, $u$
        may be monotone but may have an infinite derivative at this particular
        point, and $\pi$ may have a cusp-like shape (see panels \ref{fig}(a) and
        \ref{fig}(b));
        \item for $\alpha\in \left[0,v\frac{\sqrt{1+v}}{1-v}\right[$, $u_{Z}$
        may change sign twice and may be infinite at two singular points, whereas $\pi_{Z}$ may change
        sign three times. Thus, $u$ may follow a multi-valued shape with two singular points at their derivatives
        and $\pi$ may have a multi-valued profile especially a loop-like shape (see panels \ref{fig}(c) and \ref{fig}(d));
        \item finally, for $\alpha\in \left]v\frac{\sqrt{1+v}}{1-v},\infty\right[$, $u_{Z}$
        may never change sign and may never take infinite values. $\pi_{Z}$ may
        change sign once and may always be finite. Thus, $u$ may have a kink-like shape, whilst $\pi$ may have a single-valued profile especially a hump-like
        shape (see panels \ref{fig}(e) and \ref{fig}(f)).
    \end{itemize}
    We give some illustrations of the previous discussions. Thus,
    we may take a velocity $v=0.24$ to plot the different profiles. The aforementioned shapes are clearly
    depicted in \fref{fig}, at initial time $\tau=0$. Particularly, for the cusp-shape, the dissipative parameter is given by $\alpha= 0.351648275547$.

\end{enumerate}
In conclusion, the studies of the novel (1+1)-dimensional NLE
equation \eref{eq11} including the Vakhnenko and the variant SWSP
equations \eref{eq3} and \eref{eqq9}, respectively, may have some
scientific interests both from the viewpoint of the investigation
of the propagation of high-frequency perturbations and from the
viewpoint of the existence of stable wave formations. Thus,
applications may be found in soliton theory, geodynamics,
hydrodynamics and nonlinear optics, just to name a few.

\end{document}